\documentclass[12pt]{iopart}

\usepackage{graphicx} 
\usepackage{bm}% bold math

\begin{document}
\title[Synchronization and chaos in a pair of coupled STNOs]{Synchronization and chaos in spin-transfer-torque nano-oscillators coupled via a high speed Op Amp}

\author{C Sanid\ and\ S Murugesh}

\address{
Department of Physics, Indian Institute of Space Science and Technology, Thiruvananthapuram 695547, India}
\ead{\mailto{sanid@iist.ac.in}, \mailto{murugesh@iist.ac.in}}
\begin{abstract}
We propose a system of two coupled spin-torque nano-oscillators (STNOs), one driver and another response, and demonstrate {using numerical studies} the synchronization of the response system to the frequency of the driver system. To this end we use a high speed operational amplifier in the form of a voltage follower which essentially isolates the drive system from the response system. We find the occurrence of 1:1 as well as 2:1 synchronization in the system, wherein the oscillators show limit cycle dynamics. An increase in power output is noticed when the two oscillators are locked in 1:1 synchronization. Moreover in the crossover region between these two synchronization dynamics we show the existence of chaotic dynamics in the slave system. The coupled dynamics under periodic forcing, using a small ac input current in addition to that of the dc part, is also studied. The slave oscillator is seen to retain its qualitative identity in the parameter space in spite of being fed in, at times, a chaotic signal. Such electrically coupled STNOs will be highly useful in fabricating commercial spin-valve oscillators with high power output, when integrated with other spintronic devices.
\end{abstract}

\pacs{05.45.Xt, 75.76.+j, 75.78.-n, 05.45.Pq}
\submitto{\JPD}

\maketitle

\section{Introduction}
Extensive theoretical and experimental studies on spin-valve geometries following the discovery of spin transfer torques in magnetic multilayer structures\cite{slonc:1996,berger:1996,stiles:2006,wolf:2006} unmasked two important phenomena relevant to spintronics industry---current induced magnetization switching and self-sustained microwave oscillations in nanopillar devices\cite{myers:1999,gro:2001,kiselev:2003,rippard:2004,berkov:2008}. These were observed in $F_{1}/NM/F_{2}$ standard trilayers in which $F_{1}$ is the ferromagnetic pinned layer, which spin polarizes the input current, and $F_{2}$ is the ferromagnetic free layer whose dynamics is studied in most of the cases. $NM$ is a non-magnetic spacer layer. The self-sustained oscillations in nano-pillar devices can be understood in terms of the balance between the torque generated by the damping forces and the spin transfer torque which acts in opposite direction to the former. These spin-torque nano-oscillators (STNOs), whose oscillations are in microwave range (frequency in GHz), are excellent candidates for oscillators to be integrated into a spintronics motivated architecture. But their appeal is marred by the feeble output power from a single oscillator. 

One way of improving the output power is to synchronize several such non-linear spin torque oscillators. Two different schemes of synchronizing the STNOs are often considered. In an experiment using electrical nano-contacts at close proximity on the same mesa, Kaka et.al. \cite{kaka:2005} showed that a direct spin-wave coupling can synchronize two STNOs. This scheme has proven to be very fruitful and is replicated in various experiments\cite{manc:2005,puf:2006}. Recently attempts have been made to theoretically explain the spin wave induced coupling, predominantly using linear spin wave theory\cite{rez:2007,chen:2009}. Another effective coupling scheme uses electrically connected STNOs to get them phase locked to the ac generated by themselves. Following the experimental demonstration of injection locking of STNOs to applied ac current by Rippard et. al.\cite{ripp:2005}, it was numerically shown that an array of oscillators electrically connected in series mutually synchronize in frequency as well as in phase\cite{fert:2006}. The coupling was due to the microwave component of the common current flowing through the oscillators. This and similar coupling schemes have been explored extensively in the literature ever since\cite{akerman:2007,georg:2008,vasil:2009,zhou:2009,uraz:2010,duss:2011,dong:2012}. This way of augmenting power by an array of electrically connected phase coherent oscillators, once realized, may prove to be a great milestone towards a nano scale oscillator with useful power output. Analytical as well as numerical studies of the synchronization effects in STNOs subjected to microwave magnetic fields also appear in the literature\cite{bonin:2010,subash:2013}. We propose a novel way of electrically coupling STNOs, in a drive-response scenario, which we believe will be of substantial interest in the background of aforementioned developments.

In this work we study the various types of synchronization as well as chaotic dynamics a drive-response coupling of two STNOs can bring about. To this end, we propose a coupling using a high speed operational amplifier (Op Amp), which acts like a voltage follower. It essentially insulates the driver (master) oscillator from any feedback from the response (slave) system. The intention here is to study the dynamical response of a slave STNO to the signal input from another identical element whose dynamical behavior can be controlled. The current and applied field values fed in to the STNOs are such that they exhibit limit cycle behaviour. The oscillations can be large amplitude In-Plane (IP) oscillations (symmetric about the in-plane easy axis), or Out-of-Plane (OOP) where the precession is confined to only one of the hemispheres depending upon the initial condition. The signal generated across STNO1 by virtue of GMR effect is fed to STNO2 via the high speed Op Amp. The master-slave setup as well as the nature of coupling (which can be fine tuned using a coupling resistance, $R_{C}$, in the slave circuit) makes them a unique system not studied thus far. The time varying signal fed from the master effectively raises the dimensionality of the slave system (without coupling, the dynamics of the free layer magnetization of the slave STNO would be confined to surface of a 2-sphere, $S^{2}$, in the monodomain approximation which is employed in this work). We expect chaotic dynamics to appear in the borderline between IP and OOP oscillations for STNO2. What is remarkable is that, as the coupling resistance $R_{c}$ is changed across this borderline we observe the emergence of phase locking and synchronous precession  as well. We elaborate the various criteria which decides whether the system will go to synchronous, asynchronous or chaotic dynamics. 

In addition, we also study the properties of this system under periodic forcing. We use a small ac input current, of frequency $\omega$, in addition to the dc part to be fed to both of the STNOs. We then study how the phase portrait of slave system changes in relation to that of the master system. These considerations would be of great importance in building a robust coupled system of STNOs for enhancing micro-wave power. 

\section{Two spin-valve pillars coupled using high speed Op Amp}
The system under consideration is a regular spin valve, consisting of a conducting layer sandwiched between two ferromagnetic layers - one pinned with magnetization along ${\textbf e}_x$, the unit vector along the $x$ direction, and the other free. Further, the free layer is also subject to a constant Oersted field also along the ${\textbf e}_x$ direction. The dynamics of the macrospin magnetization of the free layer is governed by the Landau-Lifshitz-Gilbert-Slonczewski (LLGS) equation\cite{berkov:2008}.
\begin{eqnarray}
&& \frac{\partial \textbf{m}}{\partial t}-\alpha\textbf{m}\times\frac{\partial \textbf{m}}{\partial t} = \nonumber \\
&& \ \ \ \ \ \ \ -\gamma \textbf{m} \times \left( \textbf{H}_{eff} - \beta \textbf{m} \times \textbf{e}_{x} \right) ,\label{llg}
\end{eqnarray}
where ${\bf{m}} (\equiv \lbrace m_x, m_y, m_z \rbrace)$ is the normalized magnetization vector of the free layer. The effective field consisting of an external {magnetic field ($h_{ext}$)}, anisotropy field ({both in the ${\textbf e}_x$ direction, with the thin film assumed to have a uni-axial anisotropy whose easy axis is aligned along the direction of the applied filed}), and demagnetization field perpendicular to the layer, is given by:
\begin{equation}\label{heff}
\textbf{H}_{eff} =  h_{ext} \textbf{e}_{x} + \kappa m_{x} \textbf{e}_{x} - 4\pi M_{s} m_{z} \textbf{e}_{z} .
\end{equation}
{The parameter $\beta$ is proportional to the spin current density (for a given pillar geometry, and is roughly of the order of  $200\,$Oe with typical current densities of the order of $10^8\,A/cm^2$). The rescaled applied dc current, $a_{dc}$, is same as $\beta$ in what follows which has the dimensions of field intensity, frequently expressed in literature in the cgs unit Oersted.}
{The expression for $\beta$ is\cite{bazaliy:2004}:
\begin{equation}
\beta \equiv \frac{\hbar A j}{2 M_{s} V e} g(P),
\end{equation}
where A is the area of cross section, j is the current density and V is the volume of the pinned layer. g(P) is a dimensionless function  of the degree of spin polarization of pinned layer $(0 \leq P \leq 1)$, with typical numerical value $\sim$\,0.3.}
The sample parameters appearing in (\ref{llg}) and (\ref{heff}) are given values similar to that of permalloy film. So, damping constant $\alpha= 0.02$, anisotropy constant $\kappa=45$\,Oe, demagnetization field constant $4\pi M_{s}=8400$\,Oe and the gyromagnetic ratio $\gamma=1.7 \times 10^{-7}\,Oe^{-1}\ s^{-1}$. 

\begin{figure}[h]
\centering\includegraphics[width=1.0\linewidth]{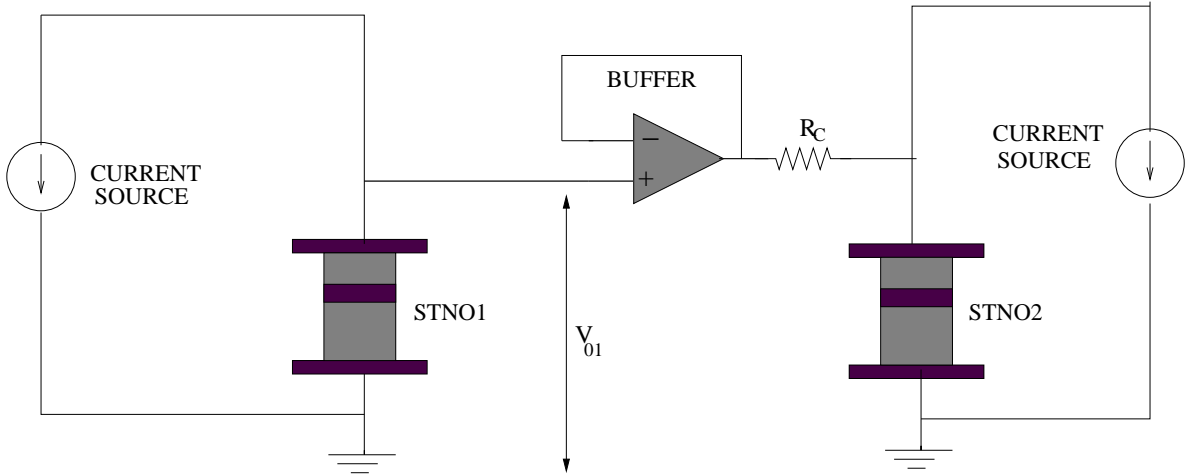}
\caption{Circuit diagram depicting the coupling using a high-speed Op Amp. The left STNO is the master and the right one is the slave, each of them separately biased using a current source. The coupling resistance, $R_{C}$ in the response circuit, turns out to be a very useful experimentally tunable parameter in this model.}
\label{circuit}
\end{figure}

We investigate the effect of coupling on the dynamical regions of the phase space of second STNO. Our coupling scheme using a high speed Op Amp is shown in figure \ref{circuit}. The Op Amp acts as voltage follower and effectively isolates the drive circuit from that of the response circuit. The voltage appearing across its non-inverting terminal is that of the STNO1 generated by virtue of GMR effect. By the property of Op Amp in buffer configuration essentially the same voltage appears across STNO2 and the coupling resistor $R_{C}$. Denoting the free-layer magnetization of STNO1 as $\textbf{m}_{1}$ and that of STNO2 as $\textbf{m}_{2}$ we derive the following pair of equations governing the dynamics of the above drive-response system:

\begin{figure}[h]
\centering\includegraphics[angle=-90,width=1.0\linewidth]{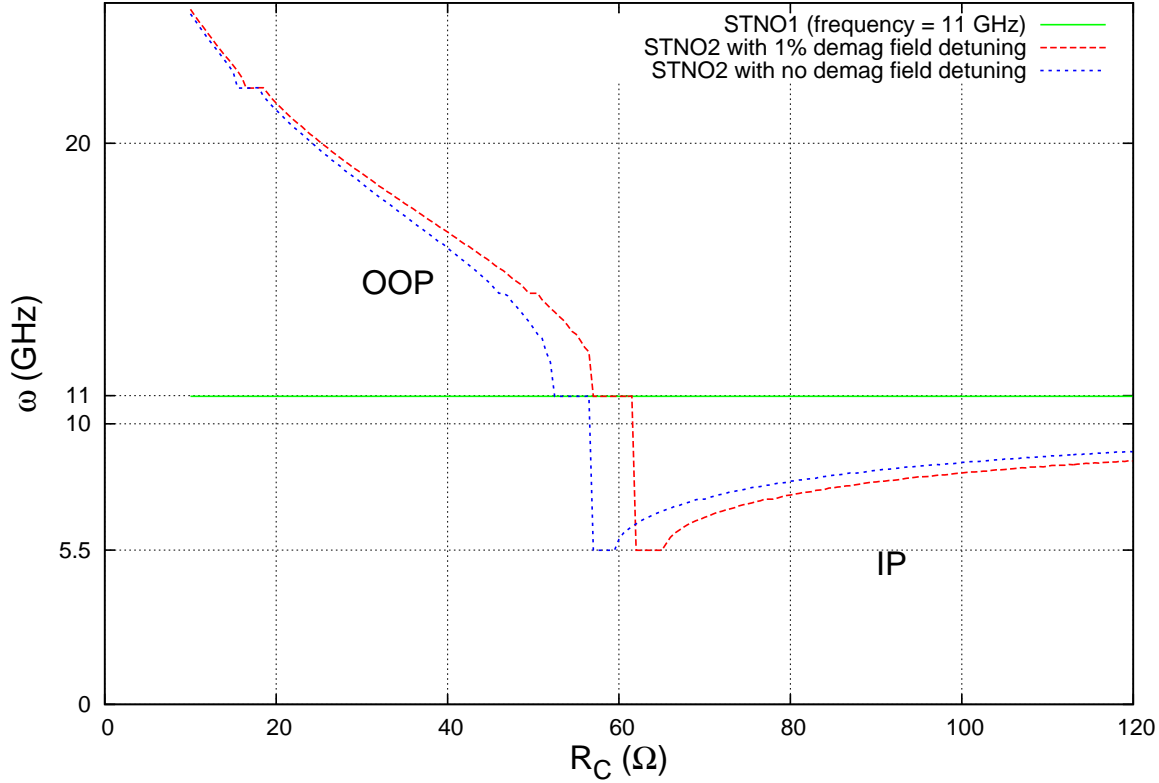}
\caption{The synchronization curve of STNO2. The parameter values are $\kappa=45$, $4\pi  M_{s}=8400$\,Oe, $R_{P}=10\,\Omega $, $R_{AP}=11\,\Omega$, $a_{dc} = 200$\,Oe and $h_{ext} = 200$\,Oe. The mismatch in the anisotropy field is 5\% and the mismatch in the demagnetization field is either 0 or 1\% as indicated in the figure. Curve flattens up to plateaus at the synchronization regime. The IP and OOP regimes of oscillations are also marked in the figure.}
\label{sync}
\end{figure}

\begin{eqnarray}\label{c1}
&&\frac{\partial \textbf{m}_{1}}{\partial t}-\alpha\textbf{m}_{1} \times \frac{\partial \textbf{m}_{1}}{\partial t} = \nonumber \\ &&\ \ \ \ -\gamma \textbf{m}_{1} \times \left( \textbf{H}_{{eff}_{1}} - \beta \textbf{m}_{1} \times \textbf{e}_{x} \right), 
\end{eqnarray}
\begin{eqnarray}\label{c2}
&&\frac{\partial \textbf{m}_{2}}{\partial t}-\alpha\textbf{m}_{2} 
\times \frac{\partial \textbf{m}_{2}}{\partial t} = \nonumber \\ 
&&\ \ \ \ -\gamma \textbf{m}_{2} \times \left( \textbf{H}_{{eff}_{2}} -  \beta ' \left( t \right) \textbf{m}_{2} \times \textbf{e}_{x} \right), 
\end{eqnarray}

where:
\begin{eqnarray}
&& \beta ' \left(t \right) = \beta \left( 1 + \frac{R_1(t)}{R_{C}+R_2(t)} \label{beta} \right) \\
&& R_{i}=R_{0}-\bigtriangleup R\ cos\left( \theta_{i}\right). \label{gmr}
\end{eqnarray}
The resistances of the two STNOs, $R_1$ and $R_2$,  depend on the dynamical state of the free layer and is modelled using the standard equation (\ref{gmr}), where $\theta$ is the angle between the free layer and the pinned layer magnetizations\cite{fert:2006}. If $R_{P}$ and $R_{AP}$ are the resistances of the spin valve in parallel and anti-parallel configurations, respectively, then $R_{0}=\left( R_{P}+R_{AP}\right)/2$ and $\bigtriangleup R = \left(R_{AP}-R_{P}\right)/2$. The right hand side of equation (\ref{beta}) comprises of contribution from coupling as well as the bias voltage of the slave STNO.

\section{Coupled dynamics - Synchronization and Chaos}

\subsection*{Synchronization}

The STNOs are given different initial conditions and are given 10\% mismatch in anisotropy field and about 1\% mismatch in demagnetization field. The coupled LLGS equation, (\ref{c1}) and (\ref{c2}), is simulated using a fourth order runge-kutta algorithm with a time step of 0.5\,ps. The inclusion of time delay (due to Op Amp action) turned out to be of no significance to the results we are presenting here and hence omitted from the discussions that follow until Section 4.

\begin{figure}[h]
\centering\includegraphics[width=1.0\linewidth]{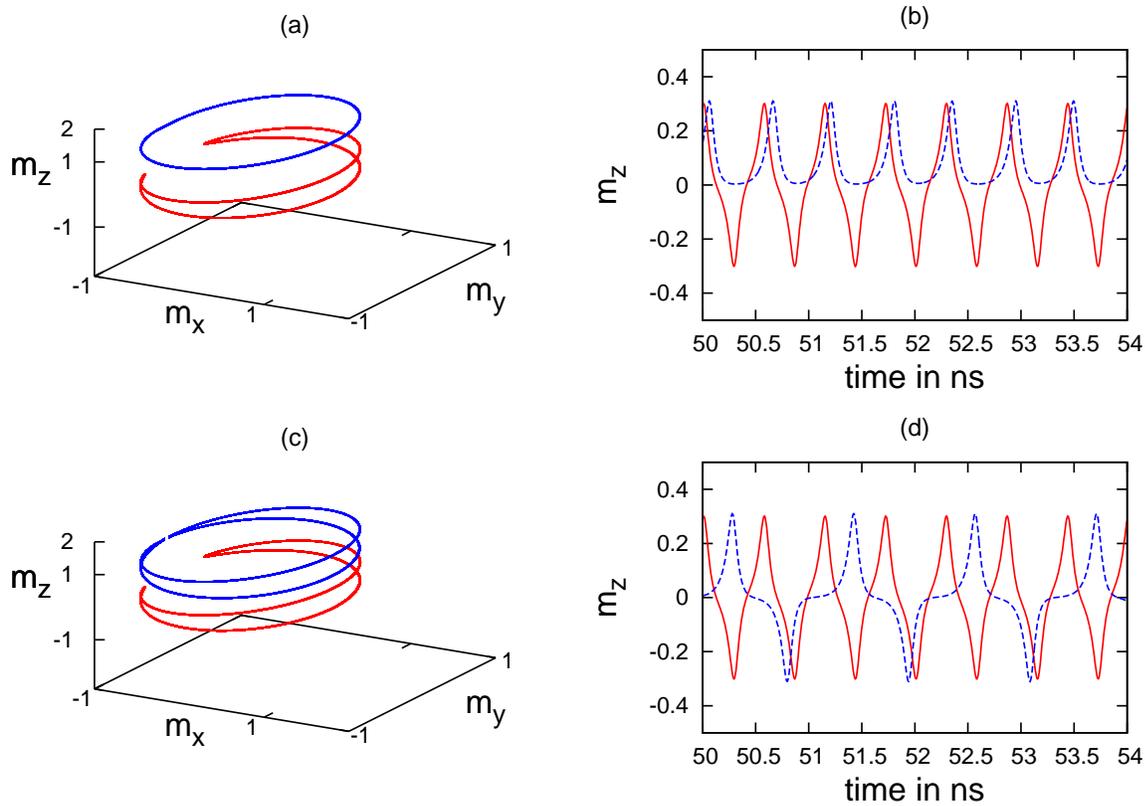}
\caption{The phase space trajectory (limit cycles) and time trace of free layer magnetization dynamics at 1:1 as well as 2:1 synchronization phases. Solid red lines (lower trajectory in (a) and (c)) denote the master where as dashed blue lines (upper trajectory in (a) and (c)) denote the slave dynamics. To avoid overlap of the figures, in (a) and (c), the trajectory of the slave oscillator (dashed blue lines) has been shifted up by 1 unit along the $m_{z}$ axis. (a) and (b) are phase space trajectory and time trace of $m_{z}$ respectively, at 1:1 synchronization region. The coupling resistance $R_{C}=60\, \Omega$ and the other parameter values are as in figure~\ref{sync}. It is clear that when the master is executing IP oscillations the slave is executing OOP oscillations. (c) and (d) are phase space trajectory and time trace of $m_{z}$ respectively, at 2:1 synchronization region. The coupling resistance $R_{C}=63\,\Omega$. It can be seen that both the master and the slave are now executing IP oscillations.}
\label{ts}
\end{figure}

When the GMR values are chosen to be $R_{P}=10\,\Omega$ and $R_{AP}=11\,\Omega$, we see the occurrence of 1:1 as well as 2:1 synchronization as plateaus in figure~\ref{sync}. {In the 1:1 synchronization regime, the master and slave STNOs precess with the same frequency, whereas in 2:1 synchronization the master STNO has double the frequency of precession as compared to slave STNO}. As the coupling resistance $R_{C}$ is increased the limit cycle frequency of the slave decreases in the OOP regime and then cross over to IP regime. After this, increasing $R_{C}$ causes the frequency to slowly go up. {This also  matches with the general response of a STNO to spin current, as increasing $R_{C}$ effectively reduces the strength of coupling\cite{bert:2005}}. Upon close inspection evidence for 1:2 synchronizations can also be found in the figure. This is discussed in some detail later in this section. The nature of free layer magnetization dynamics in these regions are further elucidated in figure~\ref{ts}. We see that there is a definite phase-locking happening between the STNOs though phase of one lags the other (figure~\ref{ts} (b) and (d)). While 1:1 mode locking, when STNO1 is undergoing IP oscillations STNO2 goes to OOP oscillation. During 2:1 mode locking both STNO1 as well as STNO2 executes IP oscillations. 

\begin{figure}[h]
\centering\includegraphics[width=1.0\linewidth]{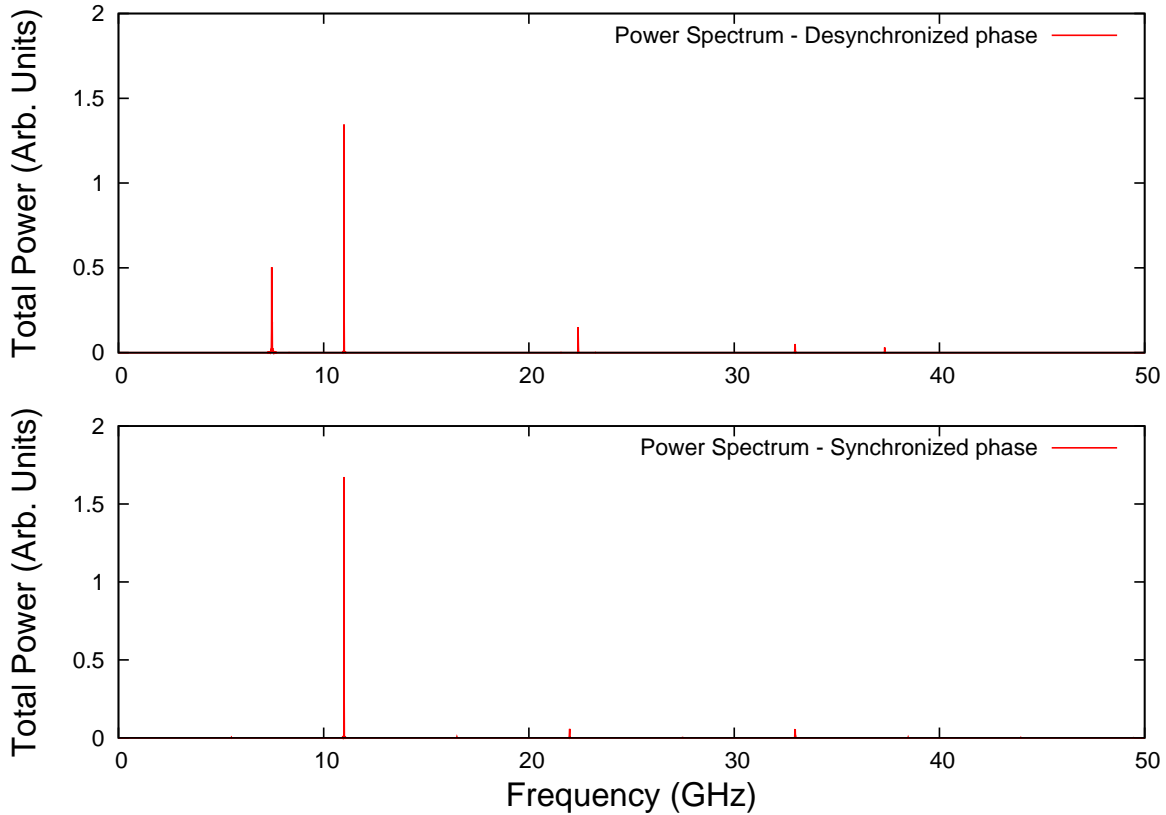}
\caption{The power spectrum for the synchronized as well as desynchronized phase. Parameter values are same as that in figure \ref{sync}. Synchronized precession is at $11$\,GHz. Desynchronized precession is at $11$\,GHz for the Master and $7.5$\,GHz for the slave. At synchronization $R_{C}=60\,\Omega$ and at desynchronization $R_{C}=80\,\Omega$. }
\label{power}
\end{figure}

\begin{figure}[h]
\centering\includegraphics[width=1.0\linewidth]{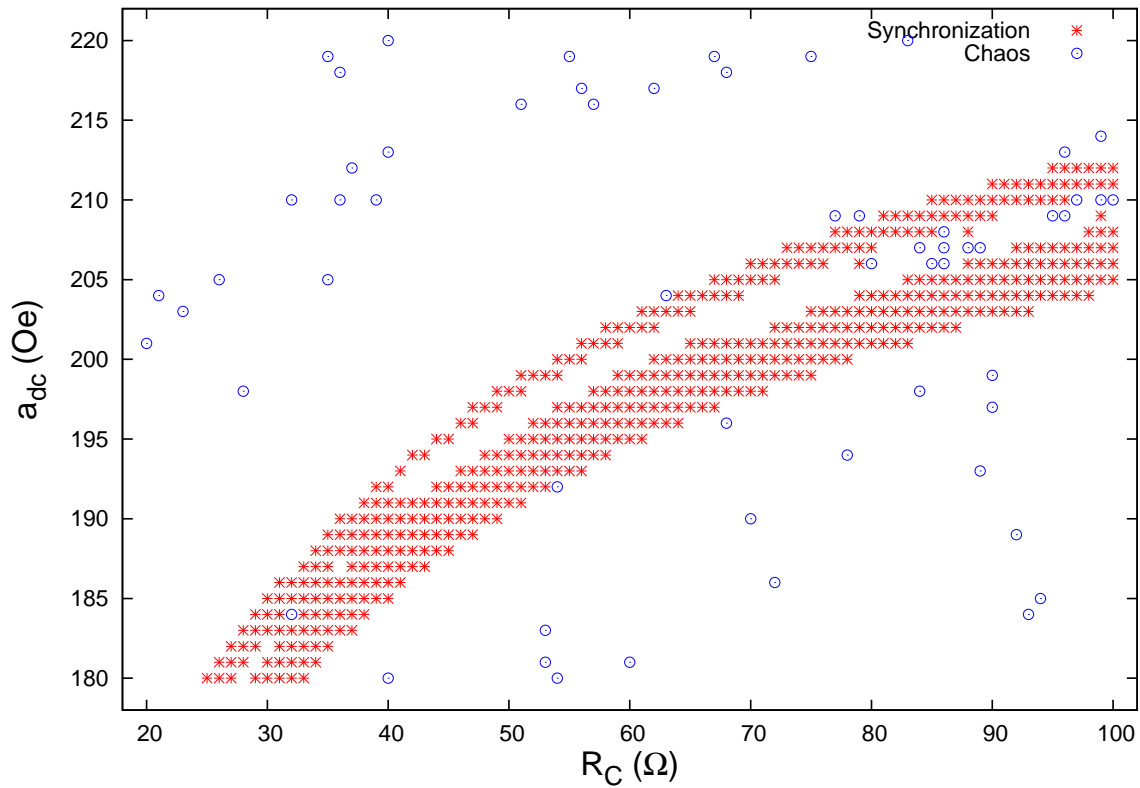}
\caption{The phase portrait in the $a_{dc} - R_{C}$ plane at the GMR value $R_{AP} = 11\,\Omega$. $h_{ext}$ is fixed at 200\,Oe. We see a well delimited synchronization region (red asterisks) surrounded by desynchronization regions (blank). Chaos is observed only at isolated points (blue circles). }
\label{phase:adc_R}
\end{figure}

In order to see the power gain at the synchronization frequency we plot the Fourier spectrum of both the STNOs in a single figure (figure~\ref{power} (b)). For comparison the scenario during desynchronization is also given at the top of the same figure. We see a distinct increase in the power at the synchronization frequency at $11$\,GHz. The power ratio of the two oscillators, an important quantity to keep track of, is found to be independent of initial condition of the slave system, a direct consequence of limit cycle motion. To further analyse the extent of synchronization we construct the phase portrait in the plane of $a_{dc}$ and $R_{c}$ which is shown in figure~\ref{phase:adc_R}. {Many points in the region} (blank) between the 1:1 and 2:1 mode locking corresponds to the multi-periodic dynamics where the dynamics jumps between the two symmetric OOP orbits but with a definite frequency. In multi-periodic case, the frequency of STNO2 differs from that of STNO1 and hence is grouped with the desynchronization region. It is evident from figure~\ref{phase:adc_R} that higher spin currents require higher coupling resistance in order to synchronize the coupled dynamics. The power ratio (between oscillator 2 and 1) remains more or less the same within the 1:1 synchronization regime, with average value 0.5 and fluctuations bounded between 0.6 and 0.4, even when the limit cycle frequency is changed by tuning the parameters. We notice that, apart from some isolated points, chaos at the boundary between IP and OOP oscillations is seldom observed at the chosen parameter values. In the next section we give a plausible explanation for the clustered chaotic points far from the synchronization region. 

\subsection*{Chaos}

\begin{figure}[h]
\centering\includegraphics[width=1.0\linewidth]{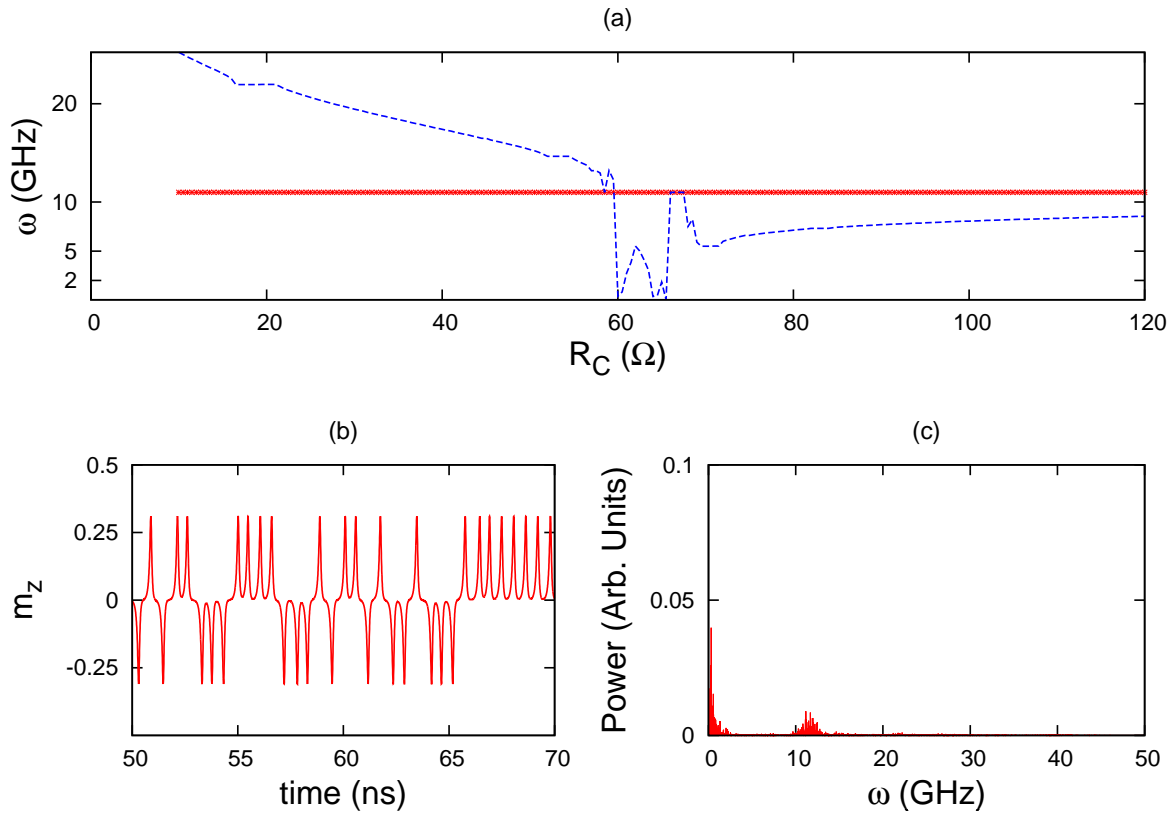}
\caption{The occurrence of chaos in coupled STNOs at the GMR value $R_{AP}=12\,\Omega$. (a)  At $R_{C}=60\,\Omega$, which showed synchronization earlier, we see the limit cycle frequency approaching zero. This is due to irregular switching of STNO2 dynamics among the available OOP and IP modes which, at these parameter values, is same as 1:1 and 2:1 synchronization modes respectively. The red line is the frequency of STNO1. (b) The time trace of $m_{z}$ displaying the random jump between IP and OOP modes. (c) The power spectrum of STNO2 showing the vanishing of the well-defined peaks. The scale of power is the same as that in figure~\ref{power}.}
\label{chaos}
\end{figure}

When the GMR values are chosen to be $R_{P}=10\,\Omega$ and $R_{AP}=12\,\Omega$, as shown in figure \ref{chaos}, we see the occurrence of chaos at the boundary between 1:1 and 2:1 synchronization regions. This is because the system switches between these modes of oscillations in a random manner. In figure~\ref{chaos} we have shown the time trace as well as the power spectrum during this phase. This is interesting because it can be used to estimate the GMR ratio itself in conjunction with other experimental techniques. During chaos, the power spectrum gets noisy and there is no useful power to be derived out of the system. Notwithstanding the commercial problems chaotic dynamics can bring about, from a dynamical systems point of view, they are still extremely important and interesting. {The effect brought about by increasing $R_{AP}$ can be understood in the following way: Increasing $R_{AP}$ essentially implies a direct increase in the GMR value which has a direct impact on the electrical coupling and can sometimes enhance the synchronization regimes\cite{fert:2006}. In our case the chaotic region seems to be sensitive to the GMR value, and more the GMR value stronger the chaotic dynamics.}

\begin{figure}[h]
\centering\includegraphics[width=1.0\linewidth]{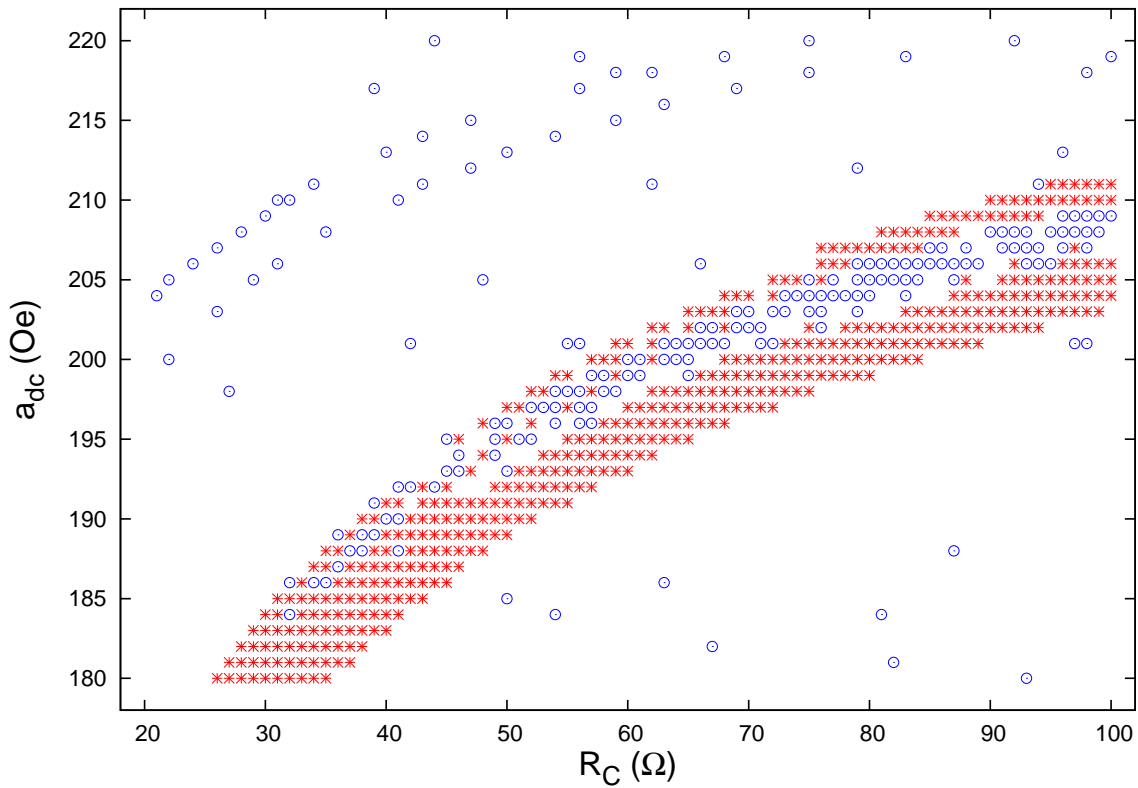}
\caption{The phase portrait in the $a_{dc} - R_{C}$ plane at the GMR value $R_{AP} = 12\,\Omega$. All other parameter values are same as in figure~\ref{phase:adc_R}. We see chaotic dynamics (blue circles) encapsulated by the synchronization regions (red asterisks). Blank regions corresponds to desynchronization dynamics.}
\label{phase:chaos}
\end{figure}

For gaining a better understanding of chaotic dynamics we turn our attention to the control space dynamics in $R_{C} - a_{dc}$ plane (figure~\ref{phase:chaos}). We see the onset of chaotic dynamics within the synchronization region itself as expected. As in the previous case, here also the dynamics turns into multi-periodic regime for some parameter values but is included in the desynchronization region in phase portraits. Thus we see that in these coupled systems where various m:n synchronizations happen in close by parameter ranges, chaotic dynamics tends to happen at the boundary between these regions. This is also crucial in noisy systems, because noise invariably make the system to randomly switch between the available states and can result in the vanishing of resonance peak even at synchronization\cite{dong:2012}. 

\begin{figure}[h]
\centering\includegraphics[width=1.0\linewidth]{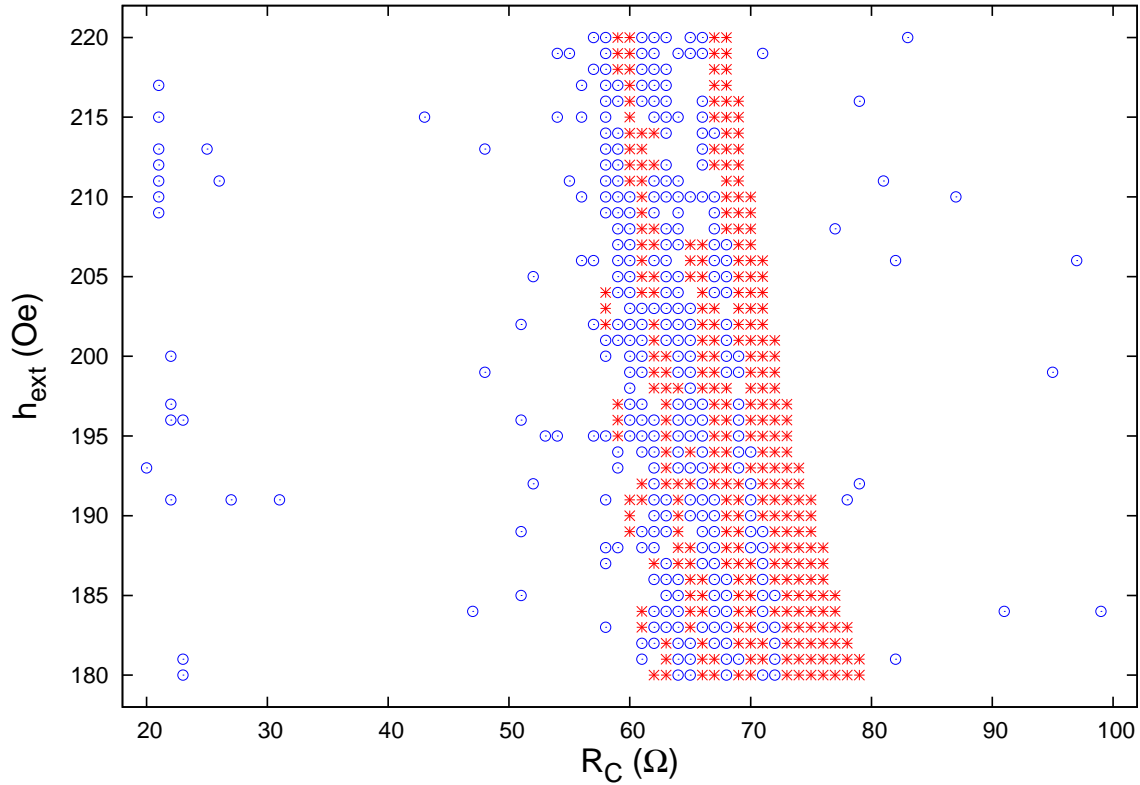}
\caption{The phase portrait in the $h_{ext} - R_{C}$ plane. $a_{dc}$ is fixed at 200\,Oe. Other parameter values and colour codings are as in figure~\ref{phase:chaos}. Here again chaos is closely tied to synchronization dynamics. }
\label{phase:chaos2}
\end{figure}

The phase picture in the $h_{ext} - R_{C}$ space also shows the embedding of chaos region within the synchronization region(figure~\ref{phase:chaos2}). Notice that chaos regions also appear outside of synchronization regions in figure~\ref{phase:chaos} as well as in figure~\ref{phase:chaos2}. This is because in the simulations we have only looked for 1:1 and 2:1 mode locking where as other m:n synchronizations are also possible in the system. We see evidence of such a locking in figure~\ref{sync}, where a small plateau appears at the frequency appropriate for 1:2 mode locking. Arguably chaotic dynamics is expected to be found associated with such higher order mode locking as well. Here it is worth pointing out that fractional synchronization in coupled STNOs are also experimentally observed\cite{uraz:2010}. 

\subsection*{Robustness under noise}
\begin{figure}[h]
\centering\includegraphics[width=1.0\linewidth]{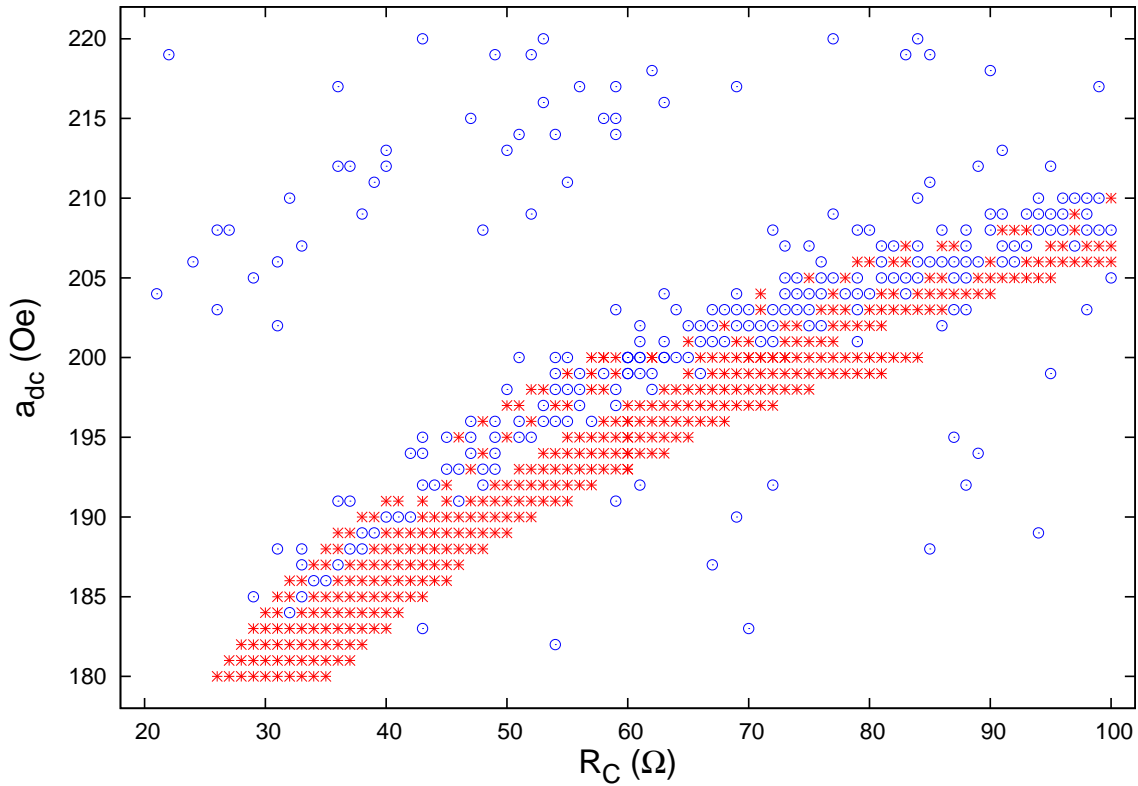}
\caption{The phase portrait in the $a_{dc} - R_{C}$ plane when a Gaussian white noise with standard deviation 0.3 was used, introducing an equivalent error of $\pm 1$\,Oe in the spin current. All parameter values and the interpretation of legends remain the same as that of figure~\ref{phase:chaos}. We see that synchronization as well as chaos regions remain more or less intact.}
\label{noise}
\end{figure}

{Real world experiments are seldom free from external noise. This can affect the reliability of our synchronization as well as chaotic regimes. In order to address the issue of robustness, we studied numerically the effect of incorporating a Gaussian white noise to the spin current, which is a good numerical approximation to thermal noise. The result of such a numerical experiment incorporating noise is shown in  figure~\ref{noise}. We notice that when a Gaussian white noise with standard deviation 0.3 was used, introducing an equivalent error of $\pm 1\,$Oe in the spin current (quite large deviation in a real experiment), our synchronization and chaotic regions remain more or less intact.}

{We even pushed the system with an error of $\pm 5$\,Oe in spin current and still found the synchronization regions intact, though more and more desynchronization regions turned to chaotic regions. We believe this suffices to state that the system under consideration is indeed robust to 
thermal fluctuations.}

\section{COUPLED DYNAMICS WITH PERIODIC FORCING}
In order to incorporate the full richness of spin-valve dynamics into our study, we let both of our STNOs to be susceptible to dynamical chaos. We use a small ac input current, of frequency $\omega$, in addition to the dc part to generate dynamical chaos. A time varying current is imperative to witness chaos in an isolated STNO, whose phase space is otherwise just two dimensional (under the macrospin assumption). Such a system displays three distinct dynamical regimes, namely \textit{Synchronization, Modifications and Chaos} in the $a_{dc}-\omega$ parameter space \cite{charles:2006}. Qualitatively, similar dynamical behavior is noticed even with a periodically alternating Oersted field instead of the alternating spin current \cite{sm:2009}.
The figure~\ref{sync} in Section 2 is applicable here with the modification that {apart from the dc biasing voltage both the STNOs are driven by ac current sources with tunable frequency as well}. We have a small ac current, in addition to dc current, flowing through both of the STNOs.

{It should be noted that this scenario is qualitatively different from the previous case in various important aspects. Here the master and slave oscillators are driven using a periodic signal, whereas in the unforced scenario only the slave STNO experiences a time varying signal (fed from the output of STNO1) in the form of coupling signal. Also, here the master STNO can go chaotic feeding the slave with a chaotic signal as shown later in this section, whereas in unforced case the slave is at best fed a periodic signal. Moreover, the meaning of synchronization itself differs considerably from the earlier case. In the unforced case, the frequency of slave STNO synchronize with that of the master STNO. In the forced case it is the synchronization of slave STNO with that of the external forcing which is considered as synchronization.}

Again the Op Amp in voltage follower mode replicates the voltage being applied to it's non-inverting terminal on it's output terminal which act as the coupling signal. In the present analysis we take in to account the time delay, $\tau$, introduced by the Op Amp action between the the two oscillators. Since this is due to the internal switching delay of Op Amp, it is taken to be a constant in the simulations ($\tau=0.05$\,ns). For the sake of numerical calculations, delay coupled oscillator pair is approximated as an array of N coupled oscillators, each having a coupling delay of $\Delta = \tau/N $ with its previous member{\cite{farmer:1982,senthil:2010}. It is noticed that time delay has no effect on the dynamics of the system and is included here for the sake of completeness of the analysis. {Our effort to introduce phase synchronization via tuning time delay has also been futile as yet. } 

The modified coupled LLGS equations are given below (see Section 2 for details):

\begin{eqnarray}\label{c1_2}
&&\frac{\partial \textbf{m}_{1}}{\partial t}-\alpha\textbf{m}_{1} \times \frac{\partial \textbf{m}_{1}}{\partial t} = \nonumber \\ &&\ \ \ \ -\gamma \textbf{m}_{1} \times \left( \textbf{H}_{{eff}_{1}} -  a\left( t \right) \textbf{m}_{1} \times \textbf{e}_{x} \right), 
\end{eqnarray}
\begin{eqnarray}\label{c2_2}
&&\frac{\partial \textbf{m}_{2}}{\partial t}-\alpha\textbf{m}_{2} 
\times \frac{\partial \textbf{m}_{2}}{\partial t} = \nonumber \\ 
&&\ \ \ \ -\gamma \textbf{m}_{2} \times \left( \textbf{H}_{{eff}_{2}} -  \beta\left( t-\tau \right) \textbf{m}_{2} \times \textbf{e}_{x} \right). 
\end{eqnarray}

where:
\begin{eqnarray}
&& a\left( t \right)=\left( a_{dc} + a_{ac}\ cos\ \omega t \right) \\
&& \beta\left(t-\tau \right) = a\left( t \right)+ \frac{a\left( t-\tau \right)\times R_1(t-\tau)}{R_{C}+R_2(t)} \label{beta_2}
\end{eqnarray}

\begin{figure}[h]
\centering\includegraphics[angle=-90,width=1.0\linewidth]{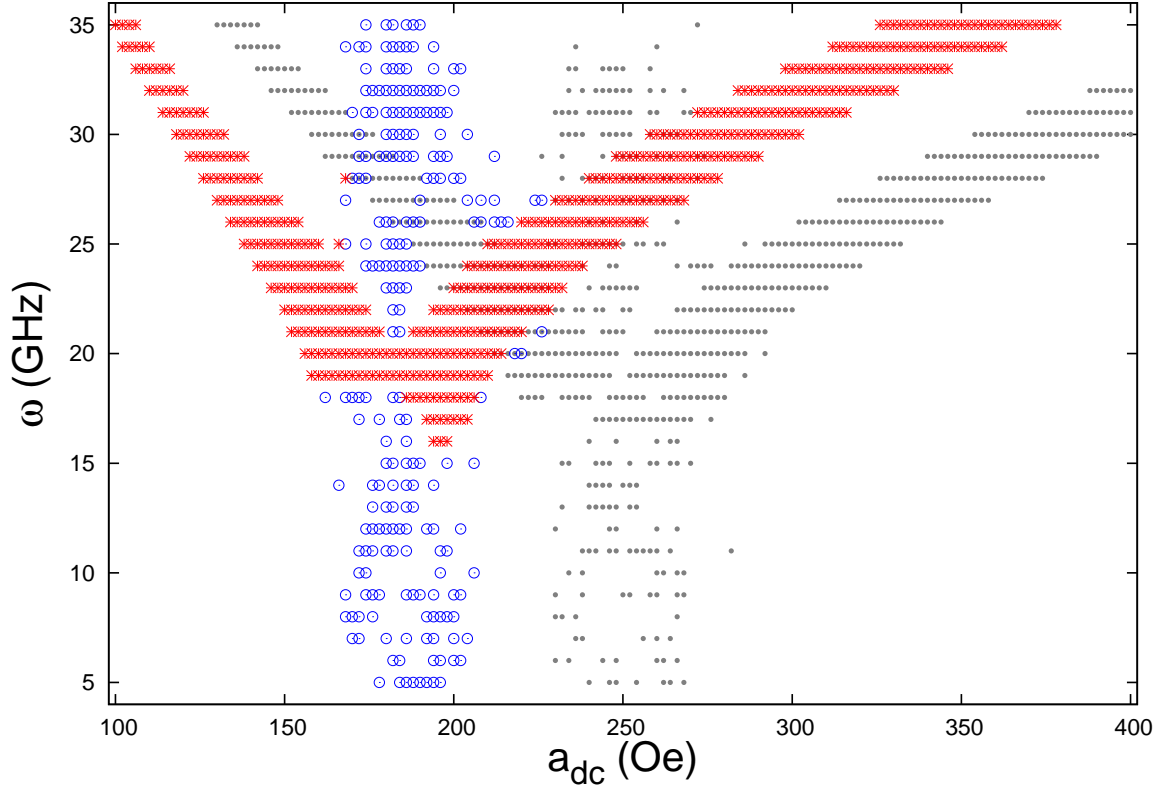}
\caption{Phase diagram of the free layer magnetization dynamics in the $a_{dc}-\omega$ plane for the slave STNO. The delay time $\tau=0.05$\,ns. The parameter values are $a_{ac}=20$\,Oe, $\kappa=0$, $4\pi  M_{s}=8400$\,Oe, $R_{P}=10\,\Omega $, $R_{AP}=11\,\Omega$, $R_{C}=20\,\Omega $. The three dynamical regions are synchronization(red asterisks), modification(blank) and chaos(blue circles). The phase diagram for the master, STNO1, shown shaded for reference, also has similar dynamic regimes.}
\label{tree}
\end{figure}

The $\omega-a_{dc}$ phase diagram for the drive system, STNO1 (figure \ref{tree}), features the \textit{synchronization branches} with a {\it chaotic stem}, as  expected (see figure 1 in \cite{charles:2006}). Interestingly, the response system, STNO2, too shows synchronization branches and a chaotic stem (red crosses and blue stars, respectively, in figure \ref{tree}) identical to that of the drive system, but with a prominent shift of the entire phase diagram towards a lower value of spin current, $a_{dc}$, with the shift determined only by the coupling resistor $R_C$. An important observation is that the qualitative picture of the phase diagram is preserved by the response STNO, in spite of being fed in, at times, a chaotic signal. One may speculate that for an extended system of N-STNOs, coupled in the manner discussed here, the individual STNOs will continue to preserve their qualitative phase (tree) structures, albeit shifted. Although the phase diagram of STNO1, the chaotic stem and synchronization branches, appears shifted compared to that of STNO2, it has to be noted that upon a careful reading the two `trees' are not exactly identical in their detail. For instance, there are points on the stem region of STNO1 that correspond to chaotic motion, but whose counterparts in the stem region of STNO2 do not.

\begin{figure}[h]
\centering\includegraphics[width=1.0\linewidth]{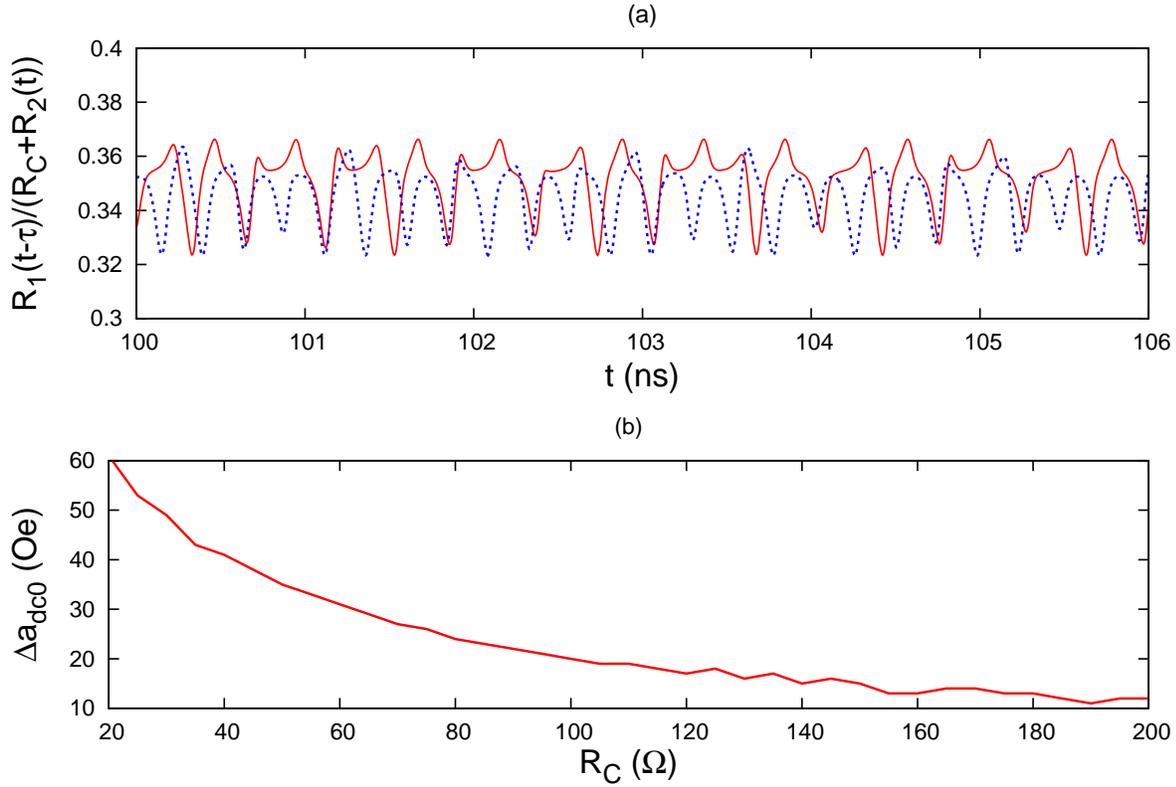}
\caption{(a) Time series of the ratio $\left( \frac{R_1(t-\tau)}{R_{C}+R_2(t)} \right) $. The average fluctuations are smaller than one but yet significant. The value of $R_{C} = 20\,\Omega$. {The red line corresponds to $a_{dc} = 250$\,Oe and $\omega = 26$\,GHz and blue lines correspond to $a_{dc} = 190$\,Oe and $\omega = 26$\,GHz}. Other parameters remains the same as that of figure~\ref{tree}. (b) The dependence of shift in critical value of current denoted as $\Delta a_{dc0}$ for the onset of chaos on the coupling resistance $R_{C}$ for $\omega = 10$\,GHz. $R_{C}$ is measured in $\Omega$s and $\Delta a_{dc0}$ in Oe. As can be seen from the figure, larger the resistance lower the shift.}
\label{shift:fluc}
\end{figure}

An important parameter in the set of coupled equations (\ref{c1_2}) and (\ref{c2_2}), is the coupling resistance in the slave circuit, $R_{C}$. For a coupling resistance of $20\,\Omega$, the shift in $a_{dc}$ is noticed to be nearly $60$\,Oe. The shift in the value of $a_{dc}$ as a function of $R_{C}$ is shown in figure~\ref{shift:fluc} (b). Agreeably, the shift in the value of $a_{dc}$ approaches zero for large values of $R_C$, when $\beta(t)$ approaches $a(t)$ and the signal from STNO1 is effectively nullified. 

We rewrite here the expression for the coefficient $\beta$, equation (\ref{beta_2}), to gain a heuristic understanding of the contribution due to coupling.
\begin{flushleft}
\begin{eqnarray}
&&\beta = a_{dc} \left( 1+\frac{R_1(t-\tau)}{R_{C}+R_2(t)} \right) + \nonumber \\
&&a_{ac} \left( cos \omega t +cos \omega (t-\tau) \frac{R_1(t-\tau)}{R_{C}+R_2(t)} \right) \nonumber \\
&&  = a'_{dc} + a_{ac}\ f(t) \label{newbeta}
\end{eqnarray}
\end{flushleft}
For some sample values of the parameters $\omega$ and $a_{dc}$ we study the temporal behavior of the term $R_1(t-\tau)/(R_C+R_2(t))$ (see figure~\ref{shift:fluc} (a)). It is noticed that this ratio shows sharp fluctuations over a period, but varies smoothly in between. For the sample values we studied, the time period of fluctuations are comparable ($\sim$0.4\,ns) to the time period of the ac part of the spin current($\sim$0.25\,ns). However, the magnitude of these fluctuations are bounded in the range of 
$0.04$, but with a significant average value compared to 1. Thus, allowing for small fluctuations, the effective value of the dc current increases ($a'_{dc}$ in (\ref{newbeta})), consequently reducing the critical value of $a_{dc}$ at which chaotic dynamics sets in. For the same reason, the time periodic part of $\beta$, $f(t)$ in (\ref{newbeta}), remains periodic with the same frequency $\omega$ as the applied spin-current. 

\medskip

\section{Discussion and Conclusion}
In summary, we have proposed a system of two coupled spin-torque nano-oscillators---a drive system and a response system---and studied it's behaviour numerically. The occurrence of 1:1 as well as 2:1 synchronization in the system are examined in detail. In the crossover region between these two synchronization dynamics we have shown the existence of chaotic dynamics and how it depends upon system parameters. We have demonstrated the power augmentation in the synchronization regimes which is of great practical importance in the current spintronics industry. We extended the study to the  coupled dynamics under periodic forcing scenario and demonstrated the interesting possibility of controlling the nature of dynamics of the response oscillator - periodic oscillations synchronized to the applied ac spin-current, or chaotic. Our simulations show a prominent shift of the chaos regions towards low spin-current side due to coupling, the shift being determined by the coupling resistor. The pivotal role played by the coupling resistor in unforced as well as forced scenarios, as an experimentally  tunable parameter for the response system, is demonstrated.
 
Commercially available ultra-high speed Op Amps (frequency $>$1\,GHz) have frequency ranges upto 2\,GHz (For example the model LMH6702 from Texas Instruments  is a 1.7\,GHz, ultra low distortion, wide band Op Amp). Though frequency of limit cycles in STNOs usually shoots above this range, making the immediate experimental realization of the coupled system impractical, we nevertheless believe higher frequency Op Amps would be available commercially in the near future. Moreover, from our results it is apparent that it is the average value of fluctuations which is responsible for dynamical effects. Hence, minor distortions in the high frequency coupling signal due to Op Amp will not alter the the results presented here.

\end{document}